\documentclass{PoS}
\def\chpt{\raise0.4ex\hbox{$\chi$}PT}
\def\schpt{S\raise0.4ex\hbox{$\chi$}PT}
\def\et{{\it et al.}}
\def\ie{{\it i.e.},\ }

\title{Electromagnetic effects on the light pseudoscalar mesons and determination of $m_u/m_d$}

\ShortTitle{Electromagnetic effect and $m_u/m_d$}

\author{S.~Basak$^a$, A.~Bazavov$^b$, C.~Bernard$^c$, C.~DeTar$^d$, E.~Freeland$^e$,
 J.~Foley$^d$, \speaker{Steven~Gottlieb}$^f$, U.M.~Heller$^g$, 
J.~Laiho$^{h}$, L.~Levkova$^{d}$\thanks{current address: Nauto, Inc., Palo Alto, CA 94301, USA},
 J.~Osborn$^i$, R.L.~Sugar$^j$, A.~Torok$^f$\thanks{current address: Intel Corporation, Hillsboro, OR  97124, USA}, D.~Toussaint$^k$, R.S.~Van~de~Water$^{l}$, R.~Zhou$^l$ \\

$^a$ NISER, Bhubaneswar, Orissa 751005, India      \\
$^b$ Department of Physics and Astronomy,
University of Iowa, Iowa City, IA 52240, USA \\
$^c$ Department of Physics, Washington University, St. Louis, MO 63130, USA\\
$^d$ Department of Physics and Astronomy, University of Utah, Salt Lake City, UT 84112, USA\\
$^e$ Liberal Arts Department, School of the Art Institute of Chicago, Chicago, IL, USA \\
$^f$ Department of Physics, Indiana University, Bloomington, IN 47405, USA\\
$^g$ American Physical Society, One Research Road, Ridge, NY 11961, USA\\
$^h$ Department of Physics, Syracuse University, Syracuse, NY  13244, USA\\
$^i$ ALCF, Argonne National Laboratory, Argonne, IL 60439, USA\\
$^j$ Physics Department, University of California, Santa Barbara, CA 93106, USA\\
$^k$ Physics Department, University of Arizona Tucson, AZ 85721, USA\\
$^l$ Theoretical Physics Department, Fermi National Accelerator Laboratory, Batavia 60510, USA

\vspace{2mm}
{\bf The MILC Collaboration}
\vspace{3mm}

E-mail: \email{sg@iu.edu}
}  

\abstract{
The MILC Collaboration has completed production running of electromagnetic 
effects on light mesons using asqtad improved staggered quarks.  
In these calculations, we use quenched photons in the noncompact
formalism.  We study four lattice spacings from $\approx\!0.12\:$fm to 
$\approx\!0.045\:$fm.
To study finite-volume effects, we used six spatial lattice
sizes $L/a=12$, 16, 20, 28, 40, and 48, at $a\!\approx\!0.12\:$fm.
We update our preliminary values for the correction to
Dashen's theorem ($\epsilon$) and the quark-mass ratio $m_u/m_d$.
}

\FullConference{The 33rd International Symposium on Lattice Field Theory\\
		14 -18 July 2015\\
		Kobe International Conference Center, Kobe, Japan}

\begin{document}

\section{Introduction}
The masses of the up, down, and strange quarks are fundamental parameters of
the Standard Model that can be determined from lattice-QCD spectrum calculations
in combination with experimentally determined masses.  These masses are
also of interest to phenomenologists.  Electromagnetic effects on the hadron
masses are often not taken into account in lattice-QCD calculations.
For a number of years the MILC collaboration has been using quenched QED
to investigate electromagnetic effects 
\cite{Basak:2008to2013}.
For the mass ratio $m_u/m_d$, it was seen that
electromagnetic contributions to the masses
of the pion and kaon were the largest source of uncertainty  
\cite{Bazavov:2009bb}.  This paper
provides an update to two conference papers from the past year \cite{Basak:2014vca,Basak:2015lla}.

To calculate $m_u/m_d$ from lattice QCD, either we need to include 
electromagnetism in the calculation and compare with the real-world masses
of pions and kaons, or we need to estimate what the pion and kaon masses would
be in a world without electromagnetism.  In other words, we need to
estimate the electromagnetic contributions to differences between charged and
neutral meson masses.  Specifically, we need 
$(M^2_{K^+} - M^2_{K^0})^\gamma$ where $\gamma$ indicates the electromagnetic
contribution.  (There is also a difference coming from the quark masses.)
Over forty years ago, Dashen \cite{Dashen:1969eg}
showed that at leading order in chiral perturbation theory
electromagnetic mass splittings for
the mesons are mass independent, {\it i.e.},
$(M^2_{K^+} - M^2_{K^0})^\gamma = (M^2_{\pi^+} - M^2_{\pi^0})^\gamma$.
Higher order effects are accounted for via a parameter $\epsilon$
defined by
$(M^2_{K^+} - M^2_{K^0})^\gamma =
(1+\epsilon)(M^2_{\pi^+} - M^2_{\pi^0})^{\rm exp}$.
This definition of $\epsilon$, adopted by the Flavor Averaging Group (FLAG)
\cite{Aoki:2013ldr} uses the experimental 
pion-mass difference on the RHS. 
In all of our prior work, we have used
$(M^2_{\pi^+} - M^2_{\pi^0})^\gamma$ on the RHS for the cental value, and
used the experimental value as an estimate of the systematic error.
Bijnens and Danielsson \cite{Bijnens:2006mk}
showed that quenched QED is sufficient for a controlled calculation of
$\epsilon$ at next-to-leading-order (NLO) in $SU(3)$ chiral perturbation theory.  Thus, our
calculation uses quenched, noncompact QED with asqtad (2+1)-flavor
QCD ensembles.

\begin{table}[tbh]
\caption{Details of the (2+1)-flavor asqtad ensembles we used.
Ensembles used only for the finite-volume analysis are denoted by an asterisk.
The quark masses $m'_l$ and $m'_s$ are the light
and strange dynamical masses used in the runs.
The  number of configurations listed  as `132+52' for the
$a\!\approx\!0.12\:$fm, $48^3\times 64$ ensemble gives values for
two independent streams, the first in single precision,
and the second in double.
Currently, we treat them as separate data, and do not average the results.
$M_\pi$ denotes the Goldstone pion mass.
\label{tab:ensembles}}
\begin{center}
\begin{small}
\begin{tabular}{llllllr}
\hline\hline
$\approx a$[fm]& Volume
& $\beta$
& $m'_l/m'_s$& \# configs.  &$L$ (fm) & $M_\pi L$ \\
\hline
0.12 & $12^3\times64^*$ & 6.76& 0.01/0.05&  1000  & 1.4 & 2.7    \\
     & $16^3\times64^*$ & 6.76& 0.01/0.05&  1303  & 1.8 & 3.6   \\
     & $20^3\times64$ & 6.76& 0.01/0.05&  2254 & 2.3 &  4.5  \\
      & $28^3\times64$ & 6.76& 0.01/0.05& \phantom{2}274 & 3.2& 6.3   \\
      & $40^3\times64^*$ & 6.76& 0.01/0.05& \phantom{2}115 & 4.6& 9.0   \\
      & $48^3\times64^*$ & 6.76& 0.01/0.05& 132+52 & 5.5 & 10.8   \\
      & $20^3\times64$ & 6.76& 0.007/0.05& 1261 & 2.3& 3.8   \\
      & $24^3\times64$ & 6.76& 0.005/0.05& 2099 & 2.7& 3.8   \\
\hline
0.09  & $28^3\times96$ &7.09& 0.0062/0.031& 1930 & 2.3&  4.1  \\
      & $40^3\times96$ & 7.08& 0.0031/0.031& 1015 & 3.3& 4.2   \\
\hline
0.06  & $48^3\times144$ & 7.47& 0.0036/0.018& \phantom{2}670 & 2.8& 4.5   \\
  & $56^3\times144$ & 7.465& 0.0025/0.018& \phantom{2}798 & 3.3& 4.4   \\
  & $64^3\times144$ & 7.46& 0.0018/0.018& \phantom{2}826 & 3.8& 4.3   \\
\hline
0.045  & $64^3\times192$ & 7.81& 0.0028/0.014& \phantom{2}801 & 2.8& 4.6   \\
\hline\hline
\end{tabular}
\end{small}
\end{center}
\vspace{-5mm}
\end{table}

We list our ensembles in Table \ref{tab:ensembles}.
The two larger volume
$a\!\approx\!0.06\:$fm and the $a\!\approx\!0.045\:$fm ensembles were not
included in our results presented at Lattice 2014 \cite{Basak:2014vca}.
The $40^3\times64$ and $48^3\times 64$ ensembles with $\beta=6.76$ were
generated after Lattice 2014 and first discussed at CCP2014 \cite{Basak:2015lla}.
The results from those two ensembles were, however, included in the 
Lattice 2014 proceedings.

\section{Finite-volume effects}
\begin{figure}
\begin{minipage}{72mm}
\includegraphics[width=72mm]{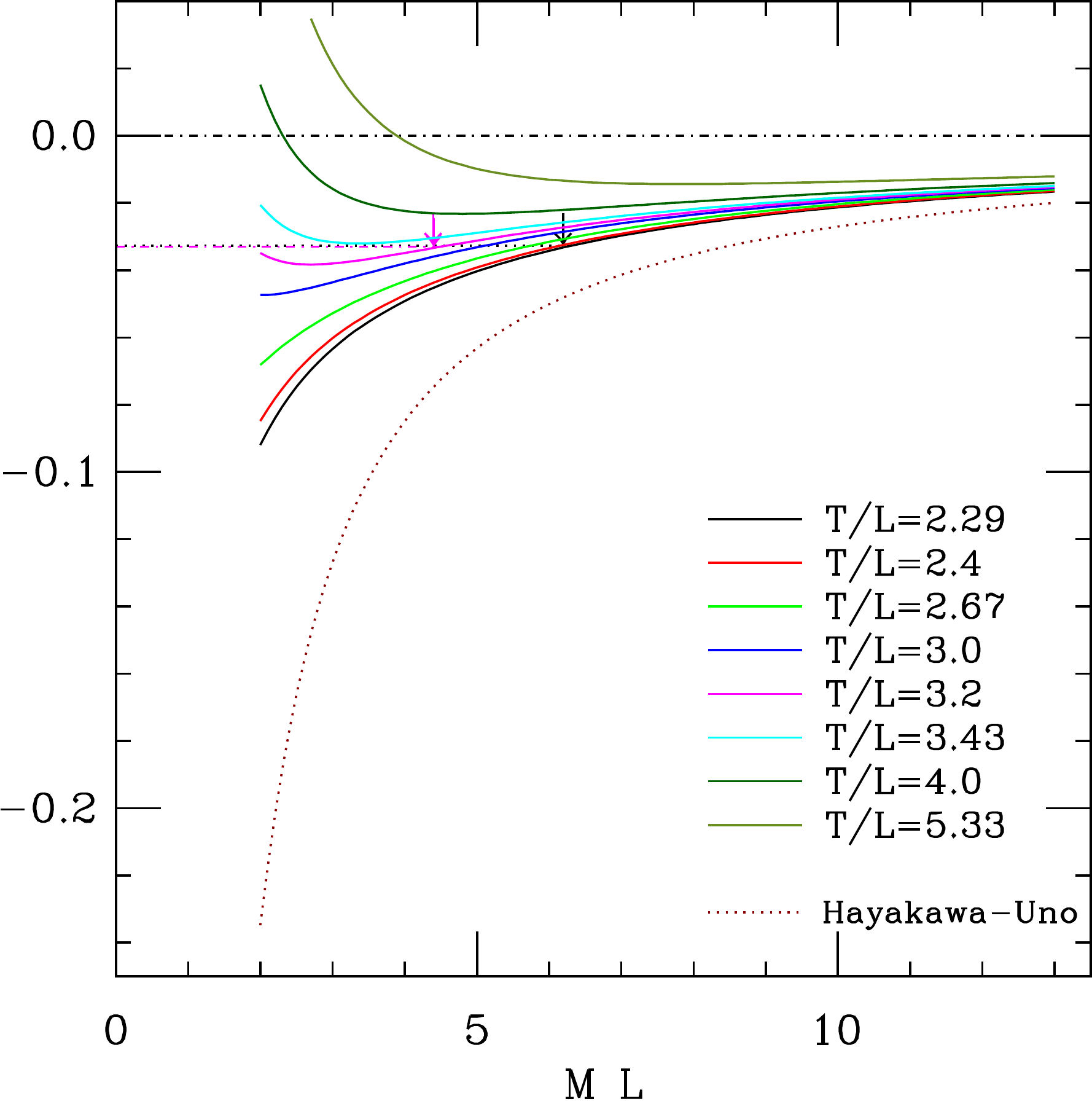}
\end{minipage}\hspace{6mm}%
\begin{minipage}{72mm}
\includegraphics[width=72mm]{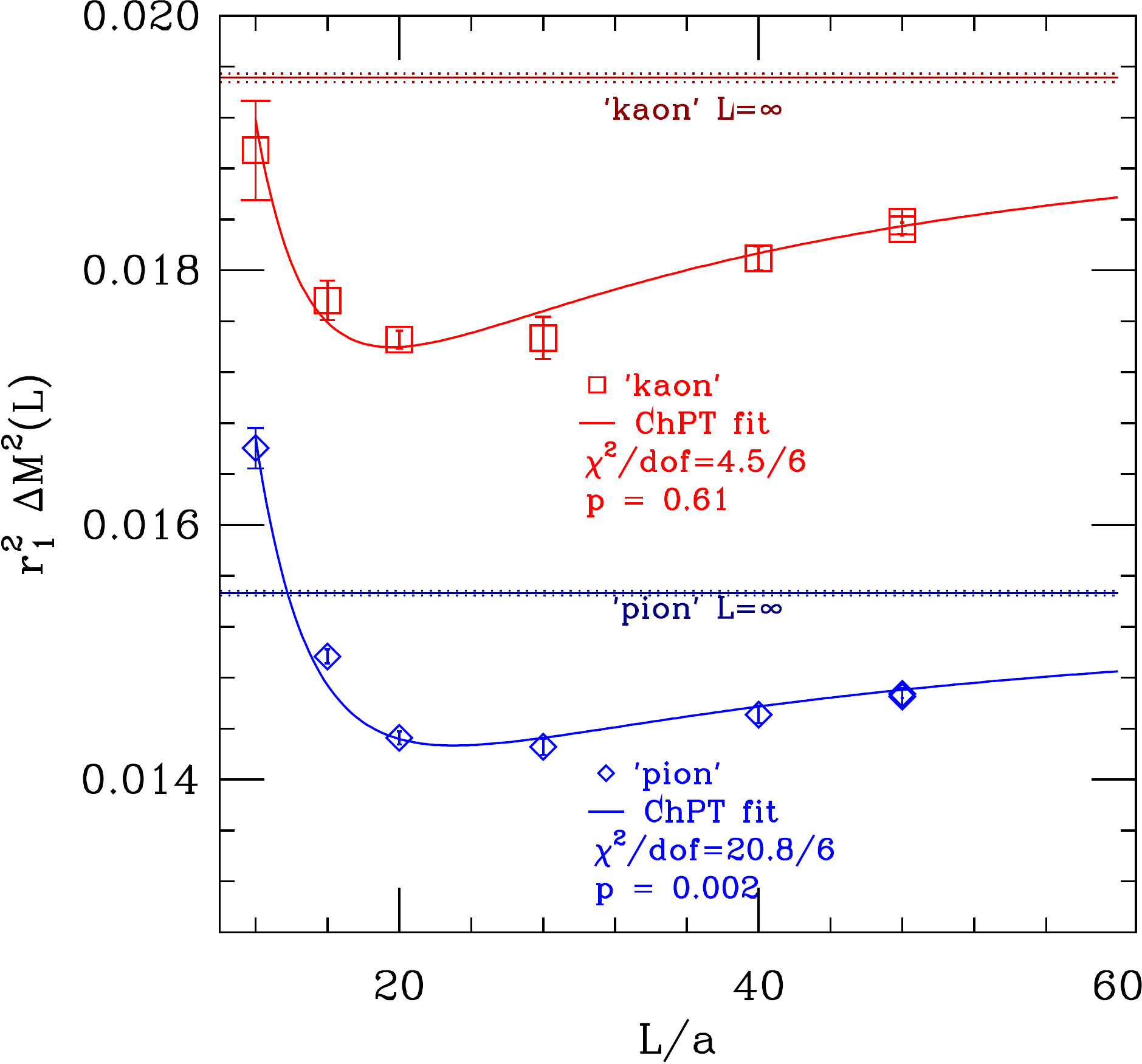}
\end{minipage}
\caption{\label{fig:MILCHU}
(L) Comparison of electromagnetic finite-volume (FV) effects in our scheme
with that in Ref.~\cite{Hayakawa:2008an}. 
The vertical axis is $(\Delta M^2(L) -\Delta M^2(L=\infty))/(q^2 M^2)$,
where $q$ and $M$ are the meson charge and mass, respectively.
The values of $T/L$ correspond to aspect ratios used in our ensembles, with
4.0 and 5.33 corresponding to the two smallest volumes in our FV
study.  Our FV effects are a factor of 2--3 smaller in most
of the relevant range.
(R) \label{fig:FV}Finite volume effects at $a\!\approx\!0.12\:$fm  and $am'_l=0.01, am'_s=0.05$ as a function of spatial lattice length $L/a$ for
two different meson masses: a unitary `pion' (blue) with degenerate masses $m_x=m_y=m'_l$, and a `kaon' (red) with masses
$m_x=m'_l$ and $am_y=0.04$, close to the physical strange quark mass.  
The fit lines are to the form from FV staggered \chpt, and have one 
free parameter each. 
The infinite-volume value are shown by horizontal solid lines with 
dotted lines for errors.
}
\end{figure}

In our earliest calculations of electromagnetic effects, we used two volumes
$20^3\times 64$ and $28^3\times64$ with a lattice spacing 
$a\!\approx\!0.12\:$fm.
Since the photon is massless, finite-volume effects need to be carefully
considered.  The uncertainty in the infinite-volume limit was a significant
source of error.  We noticed that we found small finite-volume 
effects compared to those in Ref.~\cite{Portelli:2012pn}.
Hayakawa and Uno~\cite{Hayakawa:2008an} used chiral perturbation theory
to calculate electromagnetic finite-volume effects.  The effects they found
were substantial.  However, our calculation treats the finite-volume zero
modes differently from theirs.
For the temporal modes $A_0$ in Coulomb gauge,
all modes for 3-momentum $\vec 
k=0$
must be dropped.  That is, $A_0(\vec 0, k_0) = 0$ for all $k_0$.  For the
spatial modes, the action density is
$[(\partial_0 A_i)^2 + (\partial_j A_i)^2)]/2$, so the only mode that
must be dropped is that with $(\vec 
k, k_0) = (\vec %
0, 0)$.
While we have dropped only modes with both  $\vec 
k=0$ and $k_0=0$, 
Hayakawa and Uno dropped all $A_i$ modes with $\vec 
k=0$.
The finite-size effects are smaller for our implementation of Coulomb
gauge.  However, with our method the finite-volume effect depends 
on $T/L$ where $T$ ($L$) is the temporal (spatial) extent of the lattice.
Figure \ref{fig:MILCHU}~(L) compares the finite-volume effects for the two
choices.  The figure also compares the size of the finite-volume effect for
$L/a=20$ and $L/a=28$, for a case where we were initially surprised by small
finite-volume effects.  The purple and black horizontal lines and arrows
correspond to $L/a=20$ and 28, respectively.  We see that the two lines are
very close together because they correspond to curves with different values
of $T/L$.  We also see that, with the method of Hayakawa and Uno, there is a
single dashed curve below all of ours and the intersection of vertical
lines extending down from our arrows could correspond to very different
finite-volume effects for $L/a=20$ and 28.

To test the prediction of finite-volume effects, we added ensembles
with $L/a=12$ and 16 before
Lattice 2014 \cite{Basak:2014vca}.  After the conference, we added $L/a=40$ and 48.
We show results for all six values of $L$ in Fig.~\ref{fig:FV}~(R).
In the figure, we show electromagnetic splittings of squared masses
$\Delta M^2_{xy}\equiv M^2_{xy}- M^2_{x'y'}$, where the first meson is
constructed from valence quarks  $x$ and $y$ with charges
$q_x$, $q_y$ and masses $m_x$, $m_y$, and the second meson is made
from quarks $x'$ and $y'$  with the same masses but with
the quark charges set to zero.
We multiply the squared-mass difference by $r_1^2$ where
$r_1$ is a length scale determined from the static potential.
The shape of the curve, but not its height, is determined by the 
finite-volume theory,
thus, the only free parameter in each fit is the infinite-volume limit.
The horizontal
lines show the infinite volume limit for two mass combinations labeled
`pion' and `kaon.'  It is clear why we saw such a small difference between
$L/a=20$ and 28.  The slope of each curve is low in that region of $L/a$.  
It is also worth noting that for the `pion,' the sign of the effect 
changes within the range of our new calculation.  
The astute reader may wonder why we have six degrees of freedom in our fits
when we only have six values of $L/a$.  For the largest volume, $L/a=48$, we created
ensembles using both single and double precision codes,  and fit the two
ensembles separately, so there are seven data points in the fit.
The `kaon' fit is quite good; however, the $\chi^2$ for the `pion' is rather
large mainly due to the high value for $L/a=16$.
We have greatly increased our confidence
in taking the infinite-volume limit by virtue of this test.  This is
reflected in the smaller finite-volume error we report
below for $\epsilon$ compared to previous values.
We note that the finite volume effects we have calculated are in 
agreement with those calculated by the BMW Collaboration 
\cite{Borsanyi:2014jba} for the same version of finite-volume QED 
(which they call QED$_{\rm TL}$).


\section{Chiral and continuum extrapolations}
In order to perform the chiral fit, on each ensemble we have generated 
quark propagators for a large number of valence-quark masses and 
charges.  In our early running, we sometimes included charges
with magnitude  $> 2e/3 $.
Recently, we have used charges $\pm 2e/3$, $\pm e/3$, and 0.  We then calculate
meson masses with all charge and mass combinations.  This results in many
highly correlated mass values on each ensemble, and we must thin the
data set.  We typically include 200--450 points in our fits.  We have
also experimented with uncorrelated fits.  One such fit was shown in 
Ref.~\cite{Basak:2014vca}.  Some of our fits, such as the one shown here, apply
an SVD cut to the correlation matrix to remove a few negative eigenvalues 
that are due to statistical fluctuations.

\begin{figure}[thb]
\begin{minipage}{72mm}
\includegraphics[width=72mm]{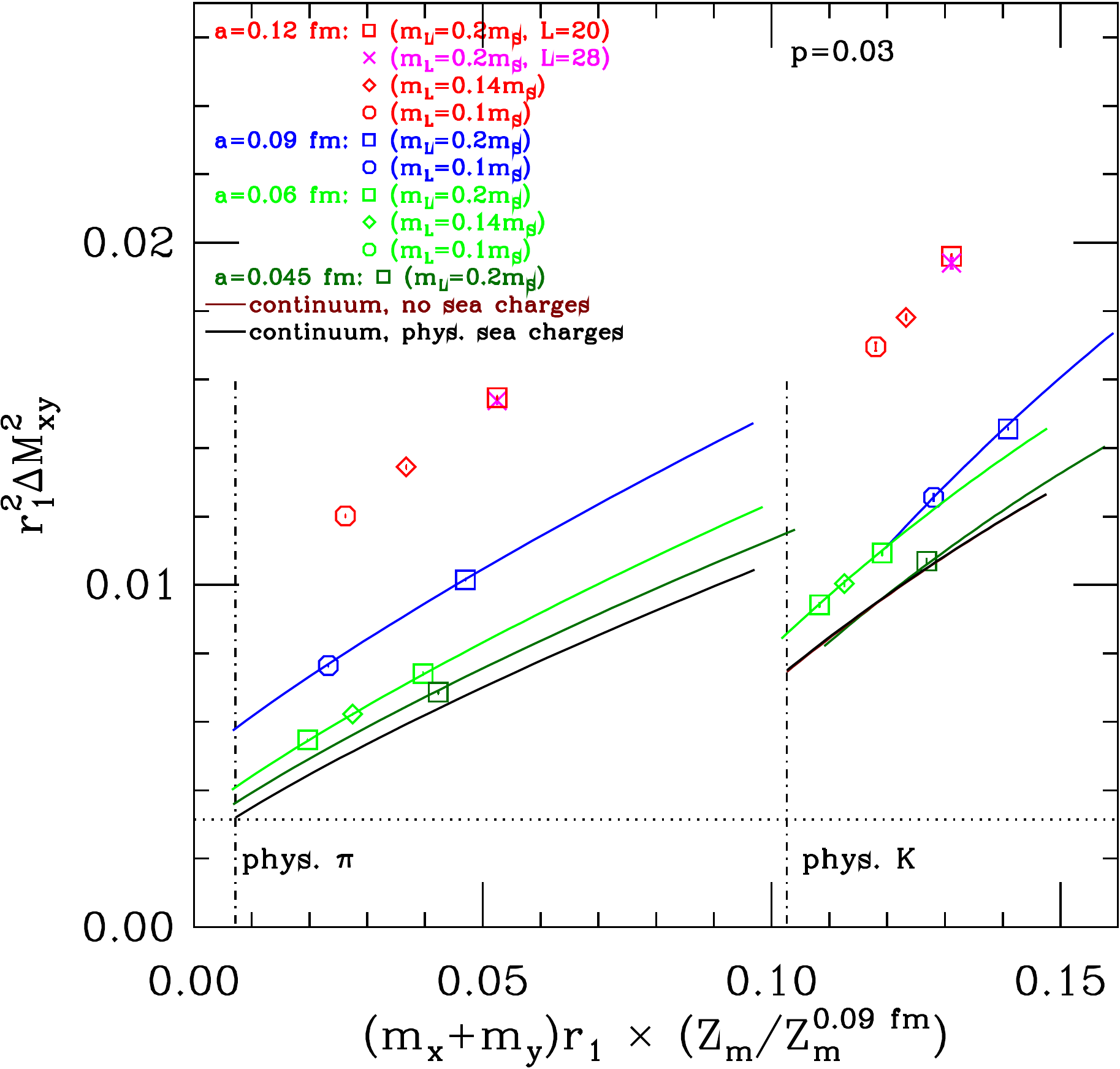}
\end{minipage}\hspace{6mm}%
\begin{minipage}{72mm}
\includegraphics[width=72mm]{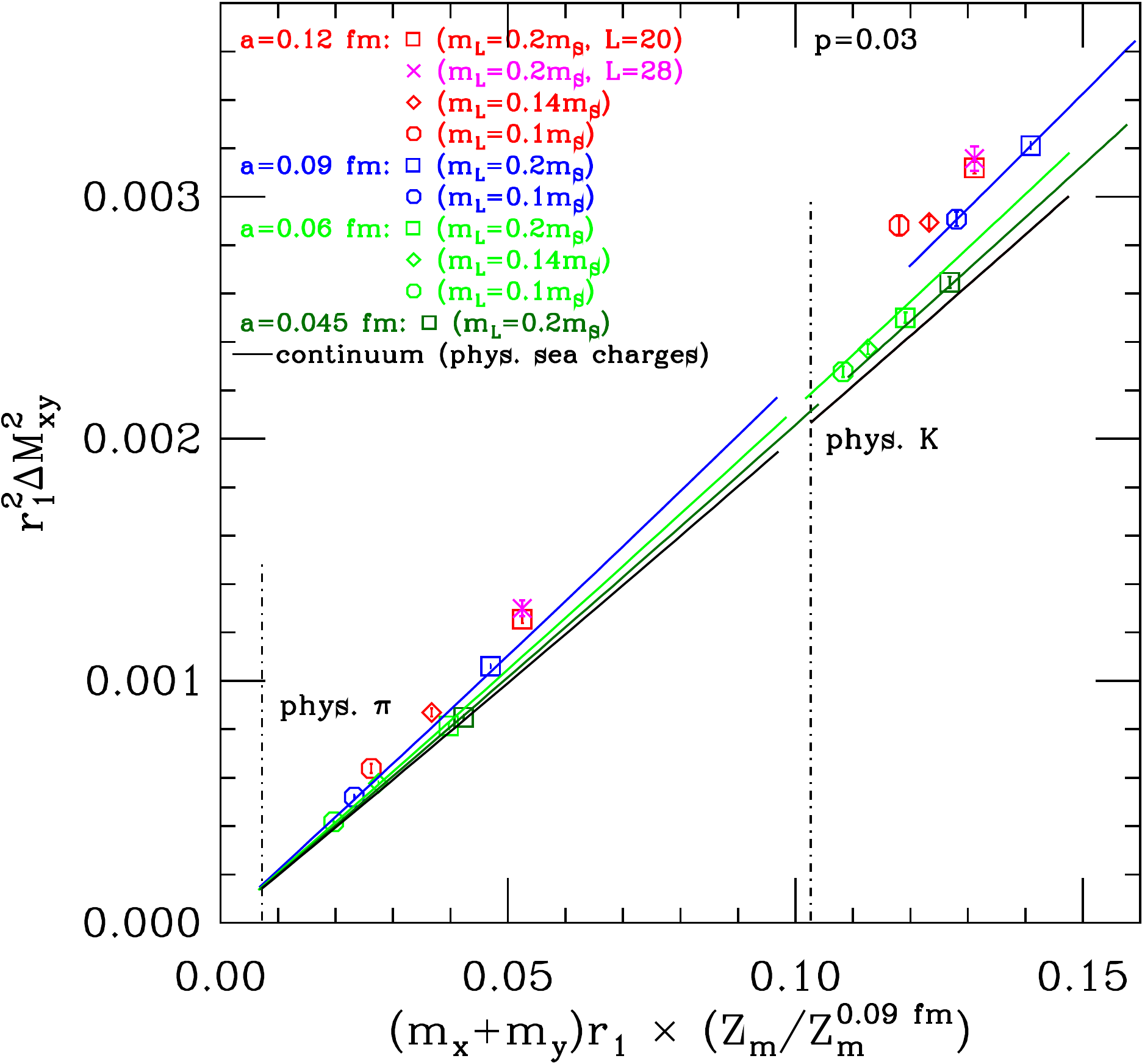}
\end{minipage}
\caption{\label{fig:chiral}Central fit to the electromagnetic splitting $\Delta M_{xy}^2$ {\it vs}.\ the
sum of the valence-quark masses.
Charge $+1$ mesons shown on left and neutral $d\bar d$-like mesons on right.
Only a small subset of the partially quenched data set included in the fit
is shown.
The data have been corrected for finite-volume effects using NLO staggered \chpt.
The blue, green, and dark green curves correspond to the three lattice spacings
included in the fit.
The vertical dashed-dotted lines show the quark-mass values for physical $\pi$ and $K$ mesons.
The black curves are extrapolations, see text.
The horizontal dotted line is the experimental value of the  $\pi^+$--$\pi^{0}$
splitting.
}
\end{figure}

In the first step of the analysis, masses are corrected for finite-volume
effects.   The corrections are 7--10\% for pions (mesons with two light
quarks) and 10--18\% for kaons (mesons with one light quark and one
whose mass is near the strange quark mass).  The corrections for kaons are
larger because of an overall factor of $M^2$ that appears in the 1-loop
diagrams.
Once the chiral fit for uncharged sea quarks is determined, we 
set valence- and sea-quark masses equal, set $m_s$ to its correct value and
take the continuum limit.  
We then use the NLO chiral logs to adjust the
meson masses to their values at the physical sea-quark charges.
The last adjustment is very small for the kaon
and vanishes identically for the pion.  Figure~\ref{fig:chiral} shows a small
subset of the data for the fit determining our central value.
An SVD cut that removes negative eigenvalues was applied.
This fit uses data from the six ensembles with $a\!\approx\!0.09\:$ fm or smaller.
A total of 444 data points are included in the fit, but ten negative
eigenvalues of the covariance matrix are removed by the SVD cut.  
As there are 37 parameters, there are 397 degrees of freedom and $\chi^2=486$.
Note how much smaller
the electromagnetic effects are for the neutral mesons (right).

Disconnected diagrams, \ie those in which the quark and anti-quark
annihilate, contribute to the physical neutral-pion propagator.  
However, such disconnected contributions are quite noisy and
hence difficult to calculate accurately.  To deal with this complication,
we drop the disconnected
diagrams and use the root-mean-squared average mass 
of $u\bar u$ and $d\bar d$ mesons.  We
call this state ``$\pi^0$.''
However, electromagnetic contributions to
neutral mesons vanish in the chiral limit, so both  the true
$(M^2_{\pi^0})^\gamma$ and our  $(M^2_{\rm{``}\pi^0\rm{''}})^\gamma$ are
small.

\begin{figure}[thb]
\begin{center}
\includegraphics[width=0.53\textwidth]{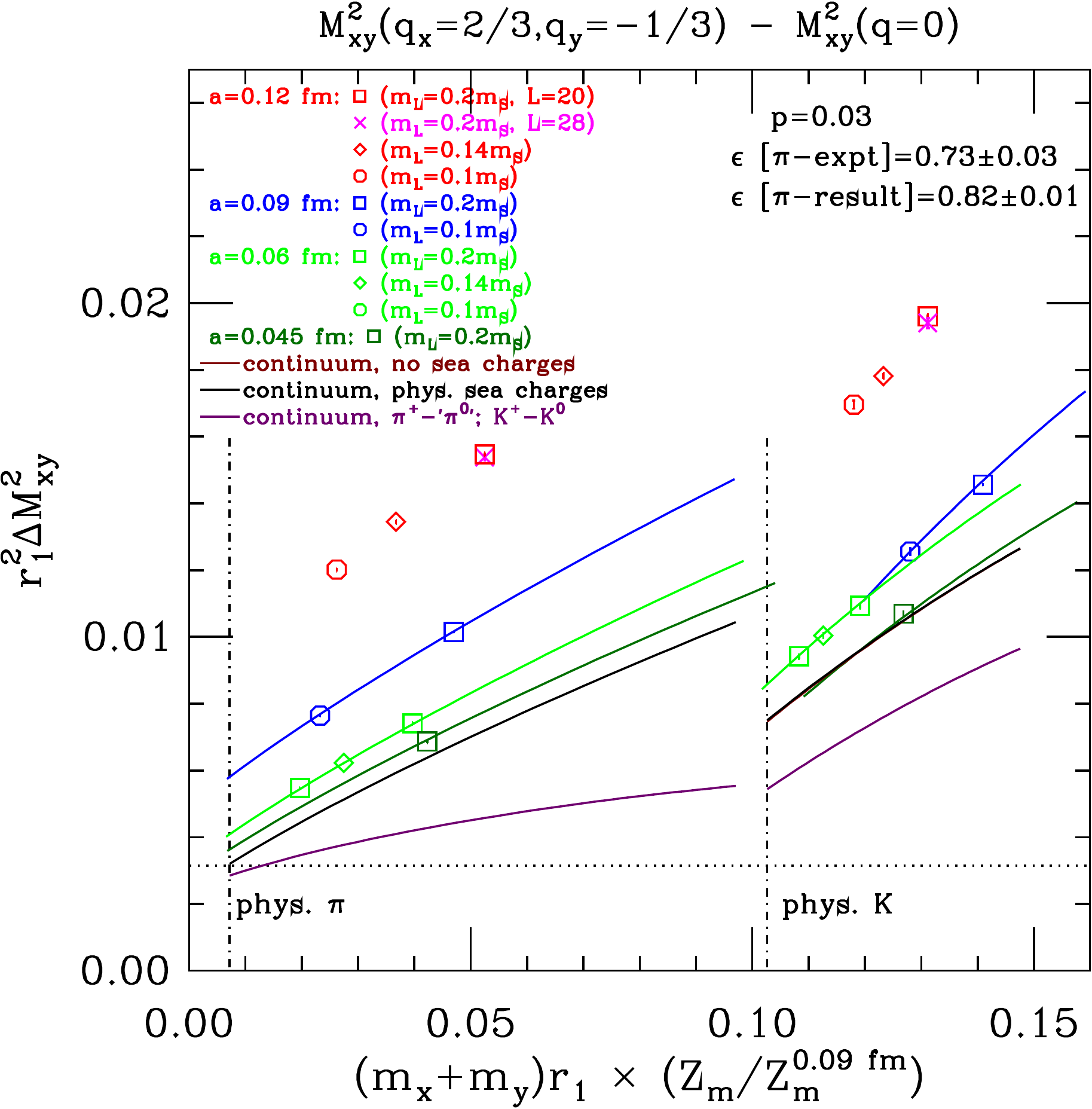}
\vspace{-5mm}
\caption{\label{fig:finalfit}Subtracting the neutral values from the charged 
values in the previous figure, we find the electromagnetic splitting $\Delta M_{xy}^2$.}
\end{center}
\end{figure}

Finally, we subtract our results for the ``$\pi^0$'' 
and $K^0$ from the corresponding charged meson to determine the physical
electromagnetic mass differences.  This is shown as the purple lines in
Fig.~\ref{fig:finalfit}.  The vertical lines represent the appropriate
sum of quark masses for the charge-averaged pion and kaon masses.
The horizontal line is the experimental value of the pion splitting.  Thus,
the distance between the horizontal line and the
intersection of the left (pion) purple line with the vertical,
dashed-dotted physical pion line is an indication of the size of the
systematic error.
Looking at the ratio of the experimental result for the pion splitting to
our kaon splitting, we get $\epsilon= 0.73(3)$.
Alternatively, we may use our result for the electromagnetic
pion splitting, and we find $\epsilon= 0.82(1)$.


\section{Results and future plans}
We consider many variations of the chiral fit, including varying
the way in which the data is thinned, whether or not
an SVD cut is used, which higher order terms are included,
and whether or not the $a\!\approx\!0.12\:$fm ensembles are included.
Our current (still preliminary) result is
\begin{equation}
\epsilon = 0.73(3)_{\rm stat}(13)_{a^2}(5)_{\rm FV}\;.
\end{equation}
The first error is statistical, the second is from variations 
in the chiral-continuum
extrapolation, and the third is 
the residual finite volume
error that may remain after our correction based on the NLO formula.
This value of $\epsilon$ is different from that presented at Lattice 2015,
as we discovered some incorrect input values in the scripts used 
to take the continuum limit.  
The central fit to the meson masses has not changed.
Adding our errors in quadrature, we find $\epsilon= 0.73(14)$.

In early work, we used the results from our asqtad ensembles to
determine $m_u/m_d$.  More recently, we have used the spectrum results
from our highly improved staggered quark (HISQ) ensembles.  These newer
results for the pseudoscalar mesons benefit from the much improved
control of the chiral extrapolation possible with the HISQ ensembles.
Using our new value of $\epsilon$ in Eq.~(4.1) with the HISQ light meson
masses \cite{Bazavov:2014wgs} gives a preliminary value for the ratio
\begin{equation}
m_u/m_d = 0.4582 (38)_{\rm stat} ({}^{+12}_{-82})_{a^2} (1)_{\rm FV_{QCD}} (110)_{\rm EM},
\vspace{-1.5mm}
\end{equation}
where here ``EM'' denotes all errors from electromagnetism, while
``FV$_{\rm QCD}$'' refers to finite-volume effects in the pure QCD calculation.
The electromagnetic error has been reduced by almost a factor of four
from our more recent papers in Ref.~\cite{Bazavov:2009bb}.

In the future, we plan to repeat the analysis of quenched 
electromagnetic effects on our HISQ ensembles, which have
several advantages for the determination of $\epsilon$.  
They have smaller discretization errors, and the
chiral extrapolation errors will be much smaller, as we have ensembles
tuned to the physical light-quark mass.  Finally, the finite-volume errors
should be reduced as the HISQ lattices are larger than those for asqtad.
We are also working on a fully dynamical $SU(3) \times U(1)$ code
to allow us to include charged sea-quark mass effects
\cite{Zhou:2014gga}.  This would enable well-controlled calculations
of many additional quantities.

{\bf Acknowledgments:}
The spectrum running was done on computers at the National Center for
Supercomputing Applications, Indiana University, the Texas Advanced
Computing Center (TACC), and the National Institute for Computational
Science (NICS).  Configurations were generated with resources provided by the
USQCD Collaboration, the Argonne Leadership Computing Facility, 
and the National Energy Research Scientific
Computing Center, which are funded by the Office of Science of the U.S.
Department of Energy; and with resources provided by the National Center 
for Atmospheric Research,  NICS, the Pittsburgh Supercomputer
Center, the San Diego Supercomputer Center, and TACC,
which are funded through the National Science Foundation's XSEDE Program; 
and by Indiana University. 
This work was supported in part by the U.S. DOE   
and the NSF. 

\end{document}